\documentclass[superscriptaddress,floatfix,showpacs,showkeys,10pt,twocolumn]{revtex4}
\usepackage{mathrsfs}
\usepackage{amssymb}
\usepackage{amsfonts}
\usepackage{amsmath}
\usepackage[dvips]{graphicx}
\usepackage{amsmath,amsfonts,amssymb,graphics,graphicx,epsfig,color,times,bbm}

\begin{document}
\title{General quantum phase estimation and calibration of a timepiece in a quantum dot system}
\author{Ping Dong}
\email{dongping9979@163.com}
\affiliation{Key Laboratory of
Opto-electronic Information Acquisition and Manipulation, Ministry
of Education, School of Physics {\&} Material Science, Anhui
University, Hefei, 230039, P R China}
\author{Zhuo-Liang Cao}
\email{zhuoliangcao@gmail.com}

\affiliation{Key Laboratory of Opto-electronic Information
Acquisition and Manipulation, Ministry of Education, School of
Physics {\&} Material Science, Anhui University, Hefei, 230039, P R
China} \affiliation{The school of science, HangZhou DianZi
University, Hangzhou, 310038, P R China}

\begin{abstract}
We present a physical scheme for implementing quantum phase
estimation via weakly coupled double quantum-dot molecules embedded
in a microcavity. During the same process of implementation, we can
also realize the calibration of timepiece based on the estimated
phase. We use the electron-hole pair states in coupled double
quantum-dot molecules to encode quantum information, where the
requirement that  two quantum dots are exactly identical is not
necessary. Our idea can also be generalized to other systems, such
as atomic, trapped ion and linear optics system.
\end{abstract}



\maketitle
\section{introduction}
Relative phase plays an important role in quantum information. The
encoding of information into the relative phase of quantum systems
has been extensively used in quantum cryptographic \cite{1}, quantum
cloning \cite{2}, geometric quantum computation \cite{3} and so on.
 Phase estimation based on discrete quantum Fourier
transform (QFT) is a comparatively good method to resolve some phase
problems. The phase estimation is a procedure of measuring an
certain unknown phase with high precision, which is also the key
ingredient for resolving some complex quantum algorithms
\cite{4a,4b,4c}, e.g. factoring problem and order-finding problem.
Therefore quantum phase estimation is a very important tool in
quantum communication and quantum computation.

\begin{figure}[tbp]
\includegraphics[scale=0.35,angle=0]{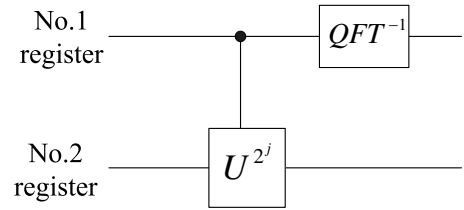}
\caption{The sketch map for the procedure of phase estimation.}
\label{fig1}
\end{figure}

In order to estimate an unknown phase $\phi$ ($\phi\in(0,2\pi])$, we
must use an oracle in the process because the phase estimation
procedure is not a complete quantum algorithm in its own right. At
the same time, the generation of a state $|u\rangle$ with an
eigenvalue $e^{i\phi}$ is necessary.  In addition, we should also
find a unitary transformation $U$, which satisfies
$$U|u\rangle=\exp({i\phi})|u\rangle.$$
Controlled unitary transformations $C-U^{2^{j}}$ ($j\in
\mathbb{N}^{+}$) will be performed in the process of the oracle
\cite{5}. The main elements of quantum phase estimation are the
oracle transformation and a inverse QFT, the sketch of which is
shown in Fig. \ref{fig1}. The No.1 register contains $m$ qubits
initially prepared in the state $|0\rangle^{\otimes m}$ while the
eigenstate $|u\rangle$ was encoded into No.2 register. The detailed
process of phase estimation can be described as following: firstly,
perform a Hadamard gate operation on each of the $m$ qubits  in No.1
register. Secondly, apply appropriate $C-U^{2^{j}}$ operations on
the whole system with the $m$ qubits in the No.1 register used as
controlled bits while $|u\rangle$ as target bit. Then apply a
inverse QFT on the qubits in No.1 register. Finally, measure the
output of No.1 register. According to the measurement result, we can
estimate the unknown phase $\phi\simeq \widetilde{\phi}$. The
successful probability and the number of digits of accuracy we wish
to have in the estimation are depend on $m$.

Recently, many researches on phase estimation have been presented
including the lower bound for phase estimation \cite{6}, optimal
phase estimation for qubits in mixed states \cite{7}, optimal phase
measurements with pure Gaussian states \cite{8} and optimal quantum
circuits for general phase estimation \cite{9}. However, the
implementation of quantum phase estimation in physical systems is
not a easy task since an unknown phase is involved in the procedure.
To overcome this difficulty, we can introduce a fungible magnitude
$T$ into the procedure of phase estimation. Solid-state system would
be the best promising candidate for quantum computer considered by
scientists. Recently one of the solid-state systems--- quantum dot
system attracts much attention because of its intrinsic properties.
In the realm of quantum dot, electronic charge states
\cite{10a,10b}, single-electron spin states \cite{11a,11b}, the spin
singlet state and triple states of double electrons \cite{12a,12b}
can all be used as qubit to encode quantum information. Especially,
schemes combining cavity technology become very useful for quantum
information processing because the cavity mode can be used as
date-bus for long-distance information transfer or long-distance
fast coupling between two arbitrary qubits. In comparison with other
transmission medium, the parallel operations on two arbitrary
different qubits can be more easily realized by using cavity
technology. Moreover the spatial separation of electronic charge
state can enhance quantum coherent \cite{13}. Therefore we
investigate the implementation of quantum phase estimation via the
interaction between weakly coupled double quantum-dot molecules and
microcavity in this paper. Because we introduce the new fungible
magnitude, we can calculate time $T$ in terms of the final
measurement result, which corresponds to the phase
$\widetilde{\phi}$. Then we can calculate the error of time
comparing with an ideal clock, if the error is within the range of
the precision $\eta$ ($\eta=\widetilde{\phi}/\phi\times100 \% $) of
phase estimation, the error will be neglected, otherwise, the
frequency of time should be regulated.

\section{implementation of phase estimation and calibration of timepiece}
In this section, we discuss a scenario for implementing quantum
phase estimation in detail. During this process, we also can check a
timepiece whether it is precise or not by comparing with an ideal
timepiece.

\subsection{Interaction between weakly coupled double quantum-dot molecule with laser fields and  microcavity}

\begin{figure}[tbp]
\includegraphics[scale=0.75,angle=0]{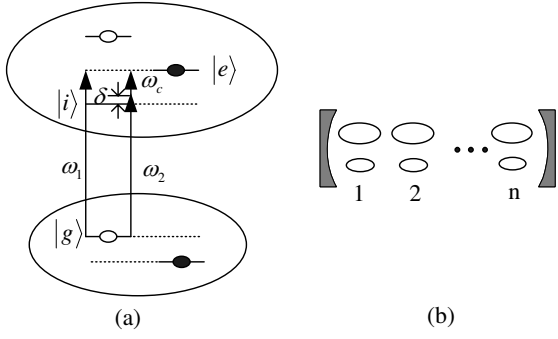}
\caption{(a) Configuration of a weakly coupled double quantum-dot
molecule. Two ellipses present two arbitrary quantum dots, the
ground state denoted by $|g\rangle$ is used for the qubit logic
state $|\widetilde{0}\rangle$, the excited state $|e\rangle$ for
logic state $|\widetilde{1}\rangle$, and $|i\rangle$ is an
intermediate state. $\omega_{1}$ and $\omega_{2}$ are two
frequencies of pulse lasers, and $\omega_{c}$ is the frequency of
cavity photon. (b) $n$ quantum-dot molecules are embedded in a
microcavity. Assume that the distance between two neighboring
quantum-dot molecules is large enough to neglect Coulomb
correlations.} \label{fig2}
\end{figure}

In our scheme, we use electronic charge (electron-hole pair) states
to store information, the configuration diagram of a qubit is shown
in Fig. \ref{fig2} (a). The state $|g\rangle$, $|e\rangle$ and
$|i\rangle$ are resulted from the conduction and valence band states
of the two individual quantum dots with different sizes
 \cite{10a}. All of the
quantum-dot molecules are embedded in a microcavity. Assume that
there is no intermediate state between the two lowest conduction and
the highest valance band state. If we perform a pulse laser on a
coupled double quantum-dot molecule with frequency $\omega_{1}$, the
Rabi transition $|g\rangle \leftrightarrow |e\rangle$ can be
governed by the following interaction Hamiltonian ($\hbar=1$)
\cite{10a}
\begin{equation}
\label{1}
H_{I}=\Omega_{1}(|e\rangle \langle g|e^{i\phi_{1}}+|g\rangle \langle e|e^{-i\phi_{1}}),
\end{equation}
where $\Omega_{1}$ is the Rabi frequency, and $\phi_{1}$ is the
laser phase.  We can obtain the evolution after a duration time $t$
\begin{subequations}
\begin{equation}
|g\rangle\rightarrow -ie^{-i\phi_{1}}\sin (\Omega_{1}t)|g\rangle+
\cos(\Omega_{1}t)|e\rangle,
\end{equation}
\begin{equation}
|e\rangle\rightarrow \cos (\Omega_{1}t)|g\rangle-i e^{i\phi_{1}}\sin
(\Omega_{1}t)|e\rangle,
\end{equation}
\end{subequations}
from which we can realize arbitrary single-qubit transformations by
adjusting $\Omega_{1}$, $t$ and $\phi_{1}$.

If we switch on a pulse laser with frequency
$\omega_{2}=E_{e}-E_{g}-\omega_{c}$, then the $\omega_{2}$ laser
photon and the $\omega_{c}$ cavity photon will participate a
resonant transition $|g\rangle \leftrightarrow|e\rangle$, the
interaction Hamiltonian can be written as \cite{10a}
\begin{equation}
\label{3}
H_{II}=\Omega_{eff}(|e\rangle \langle g|ae^{i\phi_{2}}+|g\rangle \langle e|a^{\dag}e^{-i\phi_{2}}),
\end{equation}
where $\Omega_{eff}=\Omega_{c}\Omega_{2}/\delta$;
$\delta=\omega_{2}-(E_{i}-E_{g})$ is the detuning between laser
frequency and transition energy from $|g\rangle$ to $|e\rangle$
during this transition; $\Omega_{2}$ and $\Omega_{c}$ are the
coupling strengths between $|g\rangle \leftrightarrow|i\rangle$ and
$|i\rangle \leftrightarrow|e\rangle$, respectively. There is no
occupation on the intermediate state $|i\rangle$ because of the
existing large detuning $\delta$. We can obtain the time evolution
corresponding to $H_{II}$ as
\begin{subequations}
\begin{equation}
|g\rangle|0\rangle\rightarrow|g\rangle|0\rangle,
\end{equation}
\begin{equation}
|g\rangle|1\rangle\rightarrow \cos(\Omega_{eff}t)|g\rangle|1\rangle
-ie^{i\phi_{2}}\sin(\Omega_{eff}t)|e\rangle|0\rangle,
\end{equation}
\begin{equation}
|e\rangle|0\rangle\rightarrow \cos(\Omega_{eff}t)|e\rangle|0\rangle
-ie^{-i\phi_{2}}\sin(\Omega_{eff}t)|g\rangle|1\rangle,
\end{equation}
\begin{equation}
|e\rangle|1\rangle\rightarrow|e\rangle|1\rangle.
\end{equation}
\end{subequations}
This process of evolution is the essential ingredient to realized
arbitrary two-qubit operations in this system, such as
Controlled-not gate \cite{10a} and Controlled-phase-flip, where the
photonic state ($|0\rangle$ or $|1\rangle$) is used to mediate the
coupling between arbitrary two qubits.

\subsection{Implementation of general quantum phase estimation}
To implement quantum phase estimation, we prepare two clocks (clock
1 is a precise one, the frequency of clock 2 is unknown but it's
scale is well-proportioned), a vacuum microcavity mode state
$|0\rangle$, and $m+1$ coupled double quantum-dot molecules without
excess electron in their conduction bands, where $m$ molecules are
all initialized in $|g\rangle^{\otimes
m}=|\widetilde{0}\rangle^{\otimes m}$, and the $(m+1)$th molecule is
in $|e\rangle_{m+1}=|\widetilde{1}\rangle_{m+1}$. The detailed
scenario for implementing general quantum phase estimation can be
described as the following in three steps:

(I) Firstly, perform a Hadamard gate operation on each of the
quantum-dot molecules from molecule 1 to molecule $m$, respectively,
which can be realized by the interaction as that in Eq. (\ref{1}).
Here we choose $\phi_{1}=2k\pi+\pi/2$ and
$\Omega_{1}t/\hbar=2n\pi+\pi/4$, ($k,n\in \mathbb{N}$). The time $t$
is detected by clock 1. Then we should perform a controlled phase
$C-U$ gate on quantum-dot molecules $m$ and $m+1$ (molecule $m$ is
used as control bit while molecule $m+1$ as target bit) by the
interaction as that in Eq. (\ref{3}). However, molecule $m+1$ is
remain in the state $|\widetilde{1}\rangle$ at all time, so we only
need to operate a single-qubit $\phi$ phase gate on molecule $m$ to
achieve above task ($C-U$ gate) by using the interaction as that in
Eq. (\ref{1}) by choosing $\Omega_{1}t/\hbar=2n\pi+\pi/2$ and
$\phi_{1}=\phi+\pi/2$ (the $\phi$ is unknown and can be controlled
by an unknown length $l$ of an electro-optic crystal, so it also can
be controlled by the time $T$ of going through the electro-optic
crystal). The time $T$ is detected by clock 2. Similarly, we perform
\emph{2} times $\phi$ phase transformations on molecule $m-1$,
perform \emph{4} times $\phi$ phase transformations on molecule
$m-2$, $\cdots$, and perform \emph{$2^{m-1}$} times $\phi$ phase
transformations on molecule 1. After that, the state of total system
becomes
\begin{eqnarray}
|\psi\rangle&=&\frac{1}{2^{m/2}}(|\widetilde{0}\rangle_{1}+e^{i2^{m-1}\phi}|\widetilde{1}\rangle_{1})
(|\widetilde{0}\rangle_{2}+e^{i2^{m-2}\phi}|\widetilde{1}\rangle_{2})\nonumber\\
&&\cdots (|\widetilde{0}\rangle_{m}+e^{i2^{0}\phi}|\widetilde{1}\rangle_{m})|\widetilde{1}\rangle_{m+1}\nonumber\\
&=&\frac{1}{2^{m/2}}\sum_{k=0}^{2^{m}-1}e^{i\phi k}|k\rangle.
\end{eqnarray}

(II) Setting $\phi=2\pi\varphi$, assume that $\varphi$ can be
expressed exactly in $m$ qubits, so
$\varphi=0.\varphi_{1}\cdots\varphi_{m}$ ($\varphi_{i}=0$ or $1$),
where $0.\varphi_{1}\cdots\varphi_{m}=\varphi_{1}/2+\varphi_{2}/4+
\cdots +\varphi_{m}/2^{m}$. The state of quantum-dot molecules from
molecule $1$ to  molecule $m$ can be rewritten as
\begin{eqnarray}
|\psi\rangle&=&\frac{1}{2^{m/2}}(|\widetilde{0}\rangle_{1}+e^{2\pi i0.\varphi_{m}}|\widetilde{1}\rangle_{1})
(|\widetilde{0}\rangle_{2}+e^{2\pi i0.\varphi_{m-1}\varphi_{m}}|\widetilde{1}\rangle_{2})\nonumber\\
&&\cdots (|\widetilde{0}\rangle_{m}+e^{2\pi i0.\varphi_{1} \cdots \varphi_{m}}|\widetilde{1}\rangle_{m}).
\end{eqnarray}
Then perform a inversed QFT on the No.1 register, the detailed
process can be described as following. (1) we perform a Hadamard
transform on quantum-dot molecule 1, the state of quantum-dot
molecule 1 becomes $|\varphi_{m}\rangle$. (2) we perform a series of
operations on quantum-dot molecules 1 and 2: we perform a
single-qubit $-\theta$ ($\theta=\pi/4$) phase gate operation on
quantum-dot molecule 1, a Controlled-not gate operation on
quantum-dot molecules 1 and 2 (molecule 1 is used as control bit
while molecule 2 as target bit), a single-qubit $\theta$ phase gate
operation on molecule 2, a Controlled-not gate operation on
quantum-dot molecules 1 and 2 again, and a single-qubit $-\theta$
phase gate operation on molecule 2. These operations on molecules 1
and 2 can be expressed by a total transformation
$U_{12}=U_{2}(-\theta)U_{12}(cnot)U_{2}(\theta)U_{12}(cnot)U_{1}(-\theta)$.
Then we perform a Hadamard transform on quantum-dot molecule 2. The
state of quantum-dot molecule 2 becomes $|\varphi_{m-1}\rangle$. (3)
Similarly, we apply the transformation $U_{13}$ on molecules 1 and 3
with $\theta=\pi/8$, and the transformation $U_{23}$ on molecules 2
and 3 with $\theta=\pi/4$ as the step (2). Then we perform a
Hadamard transform on quantum-dot molecule 3. The state of
quantum-dot molecule 2 becomes $|\varphi_{m-2}\rangle$; $\cdots$;
(m) We apply the transformation $U_{1m}$  on molecules 1 and $m$
with $\theta=\pi/2^{m-2}$, the transformation $U_{2m}$ on molecules
2 and $m$ with $\theta=\pi/2^{m-3}$, $\cdots$, and the
transformation $U_{m-1,m}$ on molecules $m-1$ and $m$ with
$\theta=\pi/4$ as the step (2). Finally, we perform a Hadamard
transformation on quantum-dot molecule $m$. The state of quantum-dot
molecule $m$ becomes $|\varphi_{1}\rangle$.

(III) We detect the quantum-dot molecules 1, 2, $\cdots$, $m$ by
detectors, and read out the result in reversed order. The
measurement result is
$|\varphi_{m}\varphi_{m-1}\cdots\varphi_{1}\rangle$, but readout is
$|\varphi_{1}\varphi_{2}\cdots\varphi_{m}\rangle$, so the estimated
phase $\widetilde{\phi}=\phi=2\pi0.\varphi_{1}\cdots\varphi_{m}$,
which  is precise.

\subsection{Remarks on phase estimation and calibration of timepiece}
In above process, we have assumed that $\varphi$ can be expressed
exactly in $\kappa=m$ qubits, but it is only an ideal case. For an
arbitrary value of $\varphi$, and $\kappa<m$, if we wish to
approximate $\varphi$ up to an accuracy of $1/2^{n}$, then the
successful probability should be about $1-1/(2^{m-n+1}-4)$ with
$m\geq n+1$. The unknown phase $\varphi$ can be created by
modulating the length $l=vT$ of an electro-optic crystal (such as
KDP crystal), so we can estimate the time $T$ in terms of
$\phi_{1}=2\pi\varphi+\pi/2=\varpi n^{3}_{0}vr_{63}ET/2c=\varpi
n^{2}_{0}nr_{63}ET/2$, where $\varpi$ is the frequency of electric
field, $r_{63}$ is electro-optic tensor, $v$ is the velocity of
laser through the electro-optic crystal, and $n$ and $n_{0}$ are
refractive rates for vacuum and electro-optic crystal, respectively.
In the process of implementing phase estimation, the time $T$ is
detected by clock 2, if the clock 2 undergoes $h$ scales of total
$O$ scales, we can calculate the total time by $T_{total}=OT/h$
around a circle in clock 2. In the ideal case, we compare the
$T_{total}$ with the time $T_{i}$ around a circle for the idea
clock. If $T_{total}=T_{i}$, clock 2 is an accurate one, otherwise,
the frequency of clock 2 should be regulated. In the general case,
we should first determinate the error of phase
$\eta=\widetilde{\phi}/\phi \times 100\%$, then we can calculate the
error $\eta'=T_{total}/T_{i}\times 100\%$ of clock 2. If $ \eta'\leq
\eta $, we can treat clock 2 as an accurate one, otherwise, the
frequency of clock 2 has to be regulated. In the case of
$T_{total}<T_{i}$, the frequency of clock 2 should be increased,
otherwise, the frequency should be decreased. In a word, calibrate
of timepiece includes two aspects: one is checking whether the clock
2 is precise or not, the other is if the clock 2 is not precise, we
will regulate the frequency of clock 2 according to the error of
estimated phase. Similarly, we also can estimate the length $l$ of
electro-optic crystal based on the procedure of quantum phase
estimation.

\section{discussions and conclusions}
We then discuss the feasibility of the current scheme with
experimental parameters reported in current experiments. For general
weakly coupled double quantum-dot molecule, the coupling strength
$t$ between $|e\rangle$ and $|i\rangle$ is about $0.01meV$, and the
energy difference $\Delta=E_{e}-E_{i}$ is about $10meV$, thus the
spatial separation factor
$\gamma=t^{2}/(\Delta^{2}+t^{2})\simeq10^{-6}$ \cite{10a}. In our
scheme, we use two laser pluses with different coupling strength
$\Omega_{1}$, $\Omega_{2}$, which will satisfy the condition
$\Omega_{1}\sim 10^{-3}\Omega_{2}$ according to the above value of
$\gamma$. For the process of the interaction involves two photons,
the coupling strength $\Omega_{c}$ caused by cavity field is
$300MHz$ \cite{10b,11b}, where we have assumed that
$\Omega_{2}=0.1meV$ and $\delta=1meV$ as done in Ref. \cite{10a},
resulting in $\Omega_{eff}=\Omega_{2}\Omega_{c}/\delta \simeq 30
KHz$ and $\Omega_{1}\simeq10^{-4}meV$. Therefore completion of a
single-qubit operation and a two-qubit operation will cost about
several hundreds nanosecond and $10^{-4}s$, respectively. We can
calculate the total time for completing the current scheme, which is
about $T=n(n-1)/2 \times10^{-4}s$. The coherent time of the spatial
separate charge qubits can reach tens of second \cite{13} (we can
assume $T_{c}=10s$). Comparing the time $T$ with $T_{c}$, it is
shown that the number of qubit will be $n\simeq450\gg100$ if
$T=T_{c}$, so our scheme is suitable for large-scale quantum
computation in quantum dot system.

In conclusion, we present a scenario for implementing general
quantum phase estimation via weakly coupled double quantum-dot
molecules embedded in a microcavity. The involved two quantum dots
are not necessarily to be exactly identical, which reduces the
experimental difficulty. In the same process of our implementation
of quantum phase estimation, we can also realize the calibration of
timepiece or estimation of length. The key ingredient for our scheme
is to implement the $C-U$ transformation and the reversed QFT. In
addition, the error of time (length) can be calculated by the
fidelity of quantum phase. In other words, arbitrary classical
quantity related to the estimated quantum phase can be estimated by
the same method. These classical estimation results (time $T$,
length $l$, \emph{et al}) are useful for our lives. The phase
estimation would be also an important step for fabricating quantum
computer since it is the key ingredient for complex quantum
algorithms. It also deserves to note that our idea can be
generalized to other systems, such as atom system, trapped ion
system and linear optic system.

\begin{acknowledgments}
This work is supported by the National Natural Science Foundation of
China under Grant No. 60678022, the Doctoral Fund of Ministry of
Education of China under Grant No. 20060357008, Anhui Provincial
Natural Science Foundation under Grant No. 070412060, the Key
Program of the Education Department of Anhui Province under Grant
Nos: 2006KJ070A, 2006KJ057B, KJ2007B082 and Anhui Key Laboratory of
Information Materials and Devices (Anhui University).

\end{acknowledgments}

\end{document}